\documentclass[showpacs,oneside,twocolumn,prl,amsmath,amssymb,superscriptaddress]{revtex4-1}

\usepackage{cases}
\usepackage{amsmath}
\usepackage{amssymb}
\usepackage{amsfonts}
\usepackage{amssymb}
\usepackage{dcolumn}
\usepackage{bm}
\usepackage{bbm}
\usepackage{graphicx}
\usepackage{xcolor}
\usepackage{array}
\usepackage{subfigure}
\usepackage{hyperref}

\hypersetup{colorlinks=true,
            breaklinks=true,
            pdfstartview=Fit,
            linkcolor=blue,
            citecolor=blue,
            urlcolor=blue}

\begin{document}

\title{Scalable quantum computation based on quantum actuated nuclear-spin decoherence-free qubits}

\author{Xing Rong}\thanks{These authors contributed equally to this work.}
\affiliation{Hefei National Laboratory for Physical Sciences at the Microscale and Department of Modern Physics, University of Science and Technology of China, Hefei, 230026, China}
\affiliation{Synergetic Innovation Center of Quantum Information and Quantum Physics, University of Science and Technology of China, Hefei, 230026, China}

\author{Lihong Dong}\thanks{These authors contributed equally to this work.}
\affiliation{Hefei National Laboratory for Physical Sciences at the Microscale and Department of Modern Physics, University of Science and Technology of China, Hefei, 230026, China}

\author{Jianpei Geng}
\affiliation{Hefei National Laboratory for Physical Sciences at the Microscale and Department of Modern Physics, University of Science and Technology of China, Hefei, 230026, China}

\author{Fazhan Shi}
\affiliation{Hefei National Laboratory for Physical Sciences at the Microscale and Department of Modern Physics, University of Science and Technology of China, Hefei, 230026, China}
\affiliation{Synergetic Innovation Center of Quantum Information and Quantum Physics, University of Science and Technology of China, Hefei, 230026, China}

\author{Zhaokai Li}
\affiliation{Hefei National Laboratory for Physical Sciences at the Microscale and Department of Modern Physics, University of Science and Technology of China, Hefei, 230026, China}
\affiliation{Synergetic Innovation Center of Quantum Information and Quantum Physics, University of Science and Technology of China, Hefei, 230026, China}

\author{Changkui Duan}\altaffiliation{Corresponding author: ckduan@ustc.edu.cn}
\affiliation{Hefei National Laboratory for Physical Sciences at the Microscale and Department of Modern Physics, University of Science and Technology of China, Hefei, 230026, China}
\affiliation{Synergetic Innovation Center of Quantum Information and Quantum Physics, University of Science and Technology of China, Hefei, 230026, China}

\author{Jiangfeng Du}\altaffiliation{Corresponding author: djf@ustc.edu.cn}
\affiliation{Hefei National Laboratory for Physical Sciences at the Microscale and Department of Modern Physics, University of Science and Technology of China, Hefei, 230026, China}
\affiliation{Synergetic Innovation Center of Quantum Information and Quantum Physics, University of Science and Technology of China, Hefei, 230026, China}

\begin{abstract}
We propose a novel architecture for scalable quantum computation based on quantum actuated decoherence-free (DF) qubits.
Each qubit is encoded by the DF subspace of a nuclear spin pair and has long coherence time.
A nitrogen-vacancy center in diamond is chosen as the quantum actuator to realize initialization, readout and universal control of DF qubits with fidelities higher than $99\%$.
It reduces the challenge of classical interfaces from controlling and observing complex quantum systems down to a simple quantum actuator.
Our scheme also provides a novel way to handle complex quantum systems.
\end{abstract}

\maketitle

Quantum computation has great potential to exceed the computational efficiency of any classical computers for certain tasks \cite{NielsenQCQI}, such as factoring large numbers \cite{factorlargenumber}, machine learning \cite{machinelearn} and simulation of quantum systems \cite{quantumsimulation}.
To implement the quantum computation, all the DiVincenzo criteria, i.e., a scalable physical system with well characterized qubits, initialization, long coherence times, a universal set of quantum gates and a measurement capability, must be fulfilled simultaneously \cite{DiVincenzo}.
Various physical systems have been proposed for quantum computation, ranging from superconductors \cite{superconductor}, trapped ions \cite{trappedions}, semiconductors \cite{Kane}, quantum dots \cite{quantumdots} and dopants in solids \cite{NV}, molecular magnets \cite{SMM} to hybrid systems \cite{hybridsystem}.
In these quantum computation proposals, qubits are usually classically actuated, where classical interfaces work on qubits directly.
There are several challenges for quantum computation driven by classical interfaces (upper panel of Fig.~\ref{fig1}(a)).
Firstly, if qubits are coupled to the classical interface more strongly for faster quantum gates, qubits may suffer more serious dephasing effect due to their stronger couplings to the general environment.
Secondly, for large-scale quantum computation, with the increase of qubits, the Hilbert space will grown exponentially, which dramatically increases the difficulty to control such a complex quantum system by classical interfaces.
Due to the imperfections and limitations of classical interfaces, to satisfy all the Divincenzo criteria is still very challenging, especially for large-scale quantum systems.

Here, we propose a novel solid-state architecture for a scalable quantum computation based on DF qubits and a quantum actuator.
Pairs of nuclear spins are utilized to construct DF qubits which are immune to collective noise and have long coherence time.
The quantum actuator as described by S. Lloyd \cite{quantumactuator}, is a medium that connects the quantum system of interest to the classical interfaces (see lower panel of Fig.~\ref{fig1}(a)).
All the processing of qubits, such as initializing, detecting and controlling are accomplished by the coupled quantum actuator.
This reduces the difficulty of classical interfaces from manipulating an exponential Hilbert space down to a simple quantum system.
In addition, the quantum actuator avoids the problems induced by classical interfaces, such as the voltage fluctuations of Kane's A-gate, which is one of the decoherence sources \cite{Kane}.
Obviously, it is important to find a suitable physical candidate for the quantum actuator, which requires to be well handled by classical interfaces and has appropriate couplings with qubits.
In our proposal, a nitrogen-vacancy (NV) center in diamond is used as the quantum actuator, because it has long coherence time, even approaching one second \cite{NVcohere}, and can be individually polarized, manipulated and detected with high fidelity \cite{NVreadout, tolerantgate}.
The effect of imperfection from classical control can be well suppressed by the state-of-art technology.
Some other candidates for the quantum actuator, such as SiC defects \cite{SiC}, may also be feasible in our proposal.
We show that initializing, detecting and universal controlling the DF qubits can be achieved with high fidelities.
Our approach provides a scalable solid-state architecture for quantum computation, and also put forward a novel fashion to process large and complex quantum systems.

The proposed architecture of quantum computation is shown in Fig.~\ref{fig1}(b).
Each blue solid circles stands for a DF qubit encoded in a homonuclear spin pair.
Under environment effect, a single nuclear spin state undergoes the transformation, for example, $(\mid\uparrow\rangle+\mid\downarrow\rangle)/\sqrt{2}\rightarrow (e^{i\phi}$$\mid\uparrow\rangle+e^{i\phi'}$$\mid\downarrow\rangle)/\sqrt{2}$, which will finally evolve into a mixed state due to averaging over random phase differences $\phi - \phi'$.
In contrast, since $(\mid\uparrow\downarrow\rangle \pm \mid\downarrow\uparrow\rangle)/\sqrt{2}\rightarrow (e^{i\phi}$$\mid\uparrow\rangle e^{i\phi'}$$\mid\downarrow\rangle\pm e^{i\phi'}$$\mid\downarrow\rangle e^{i\phi}$$\mid\uparrow\rangle)/\sqrt{2}= e^{i(\phi+\phi')}(\mid\uparrow\downarrow\rangle \pm \mid\downarrow\uparrow\rangle)/\sqrt{2}$, these two-spin states are invariant under collective dephasing.
Hence the collective dephasing can be eliminated within the DF subspaces spanned by $(\mid\uparrow\downarrow\rangle \pm \mid\downarrow\uparrow\rangle)/\sqrt{2}$.
The DF qubits have long coherence times because of their immunities to collective noise. 
Here, a molecule with two neighboring identical nuclear spins can be used to encode such a DF qubit.
Endohedral fullerenes and halogenated fullerenes, such as a single molecule of water encapsulated in a fullerene (H$_2$O$@$C$_{60}$) \cite{H2OC60} and a fullerene containing two fluorine atoms \cite{C60F2}, are applicable candidates.
In our model, the molecule H$_2$O$@$C$_{60}$ is taken as an example.
By utilizing the scanning probe techniques with high resolution, the nano-scaled fullerenes can be precisely positioned \cite{STM}, constructing a large number of well-ordered qubits.
An array of DF qubits can be positioned with an appropriate separation distance, i.e. 2~nm, and the direct coupling between two DF qubits is at the sub-hertz level, which is negligible.
This distance is also far enough for individually controlling target DF qubit when a quantum actuator locates above it, while close enough to entangling DF qubits via a quantum actuator located between them (see the details in supplemental materials (see details in Supplementary Material)).
Coupling between the quantum actuator and a DF qubit can lead to the transfer of states between them, which can be utilized to initialize and to measure the DF qubit.
When the quantum actuator maintains on distinct eigenstates, the designated DF qubit will evolve under different Hamiltonians.
Via engineering the actuator's states, a full controllability of a single DF qubit can be achieved.
Additionally, nontrivial two-qubit gates, i.e. a CNOT gate, can also be accomplished through an indirect interaction bridged by the quantum actuator.
Thus a universal set of quantum gates for DF qubits is achieved.
Detailed descriptions are given as follows.

\begin{figure} 
\centering
\includegraphics[width=1\columnwidth]{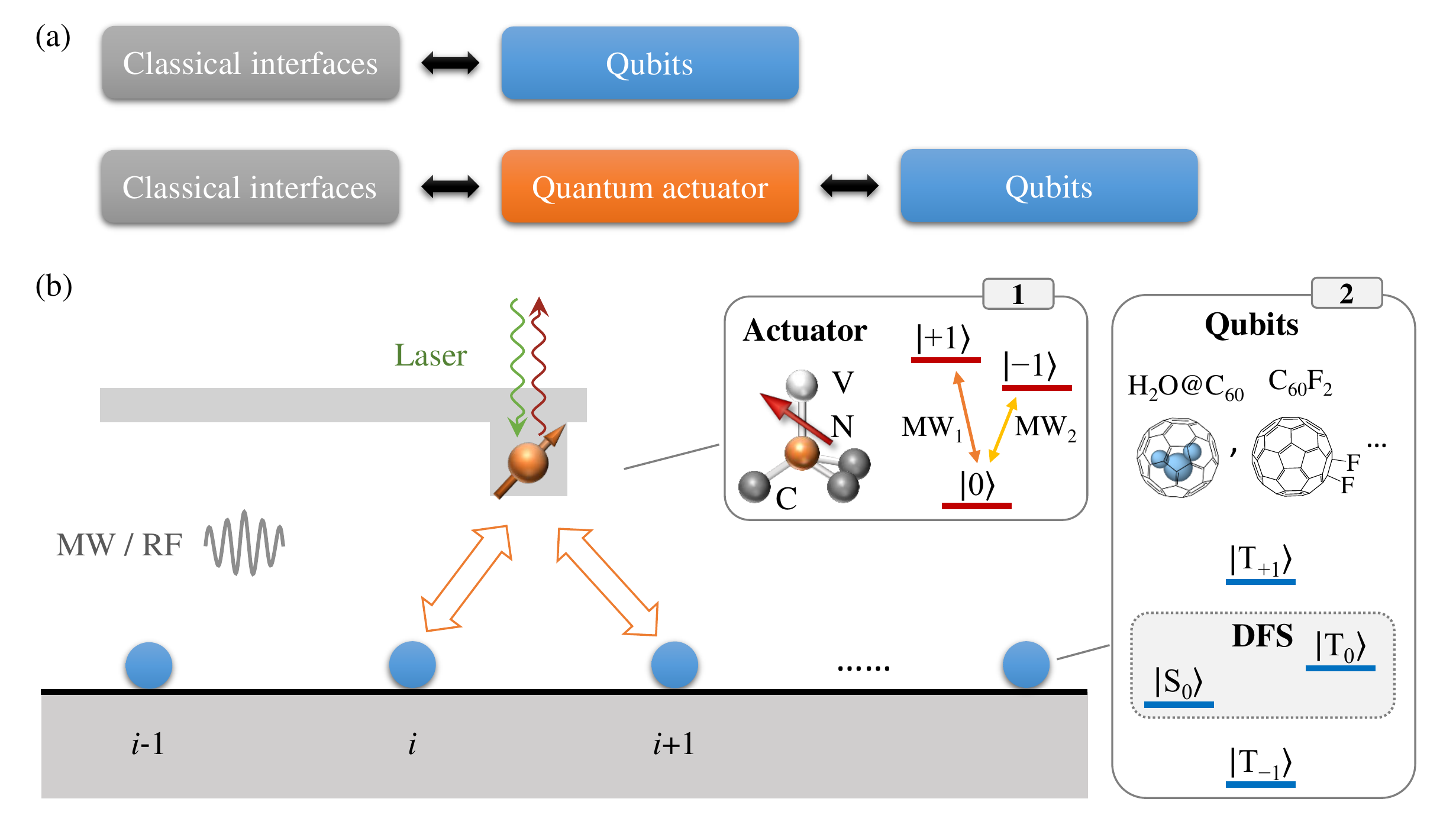}
\caption{Architecture of quantum computation. (a) Comparison of two types of architectures for quantum computation.
The upper panel shows the schematic diagram of classically actuated quantum computation, in which classical interfaces interact with the qubits directly.
The lower panel shows the schematic diagram of quantum actuated quantum computation, in which a quantum actuator is introduced to control the qubits and classical interfaces act on the quantum actuator.
(b) Detailed architecture of quantum actuated quantum computation. Ordered qubits (blue solid circles) array on the base with appropriate separations.
Inset labelled by 1 shows the atomic structure and energy levels of the NV center (the quantum actuator), which is manipulated by MW pulses (MW$_1$ and MW$_2$), and polarized/read by green/red laser pulses.
Inset labelled by 2 shows the energy levels of the nuclear spin pair and some candidate molecules for encoding DF qubits, such as H$_2$O@C$_{60}$ and $\textrm{C}_{60}\textrm{F}_2$.
Each qubit is encoded in the decoherence-free subspace (DFS) of a pair of nuclear spins.
\label{fig1}
}
\end{figure}

\textit{Hamiltonian.}
The external static magnetic field $B_0$ with strength of 500~Gauss is set along the NV symmetry axis ([1 1 1] crystal axis) and the intrinsic nitrogen nuclear spin can be polarized by laser pumping.
The polarization of the nitrogen nuclear spin can also be achieved with other field strength \cite{polarizNitrogen}.
The total Hamiltonian of the system is composed of three parts, the quantum actuator (an NV center), the nuclear spin pairs and their interactions, written as
\begin{equation}
\begin{split}
H_0= H_{\textrm{NV}}+H_{\textrm{pairs}}+H_{\textrm{NV-pairs}}.
\end{split}
\end{equation}
The Hamiltonian of the quantum actuator is given by $H_{\textrm{NV}}=DS^2_z+\omega_eS_z$, where $\textbf{S}$ is the spin-1 operator, $\omega_e$ is the Zeeman frequency corresponding to the static magnetic field $B_0$, and $D = 2.87~\textrm{GHz}$ is the zero-field splitting.
The energy levels of the quantum actuator are shown as inset labelled by 1 in Fig.~\ref{fig1}(b).
The quantum actuator can be driven with microwave (MW) pulses for the transition, e.g. $|0\rangle\leftrightarrow |\textrm{+}1\rangle$ (or alternatively $|0\rangle\leftrightarrow |\textrm{-}1\rangle$), reduced to an effective two-level system.
The ground state of this quantum actuator can be optically initialized and detected.

The Hamiltonian of nuclear spin pairs is written as $H_{\textrm{pairs}}=\sum\limits _m (\omega_iI_z^{m,p}+\omega_iI_z^{m,q}+\textsl{\textbf{I}}^{m,p} \cdot \textsl{\textbf{D}}^{m,pq} \cdot \textsl{\textbf{I}}^{m,q})$,
where $\textbf{I}^{m,p}$ and $\textbf{I}^{m,q}$ are spin-1/2 operators for spins in $m$th spin pairs, $\omega_i$ is the nuclear spin Zeeman frequency, and $\textbf{D}^{m,pq}$ is the interaction tensor between nuclear spin $P$ and $Q$ in $m$th spin pairs.
The interaction between the quantum actuator and the nuclear spin pairs is $H_{\textrm{NV-pairs}}=\sum\limits_m(\textsl{\textbf{S}}\cdot \textsl{\textbf{D}}^{m,p} \cdot \textsl{\textbf{I}}^{m,p} +\textsl{\textbf{S}}\cdot \textsl{\textbf{D}}^{m,q} \cdot \textsl{\textbf{I}}^{m,q})$,
where $\textbf{D}^{m, p(q)}$ is the interaction tensor between the quantum actuator and the nuclear spin $P(Q)$ in $m$th spin pairs.
The energy levels of the nuclear spin pair are shown as inset labelled by 2 in Fig.~\ref{fig1}(b), where $|T_{+1}\rangle=\mid\uparrow\uparrow\rangle$, $|T_{-1}\rangle=\mid\downarrow\downarrow\rangle$, $|T_0\rangle=(\mid\uparrow\downarrow\rangle + \mid\downarrow\uparrow\rangle)/\sqrt{\textrm{2}}$ and $|S_0\rangle=(\mid\uparrow\downarrow\rangle -\mid\downarrow\uparrow\rangle)/\sqrt{\textrm{2}}$.
The logical bases of DF qubits are defined by $|0\rangle_L\equiv |S_0\rangle$ and $|1 \rangle_L \equiv |T_0\rangle$.
DF subspace has been widely studied both theoretically and experimentally \cite{dfs1,dfs2,dfs3}.
Recently, the DF subspace composed of a pair of $^{13}$C nuclear spins was demonstrated in Ref.~\cite{dfs4}.
Here, we can use the regular pairs of extrinsic protons to construct DF qubits.
The DF qubits are robust to decoherence, for example, uniform static or dynamic magnetic field fluctuations can be eliminated within the DF subspace as described above, while some other noises such as the unwanted hyperfine interactions with spin bath can also be dramatically suppressed \cite{singletrelax}.
Additionally, through controlling the quantum actuator, dynamic decoupling on DF qubits can even be accomplished to resist decoherence.
Thus we can expect a longer coherence time of the DF qubits than the single nuclear spins \cite{dfs2,singletrelax,longlive} whose coherence time can reach even up to hours \cite{nucleartime1}.

\begin{figure} 
\centering
\includegraphics[width=1 \columnwidth]{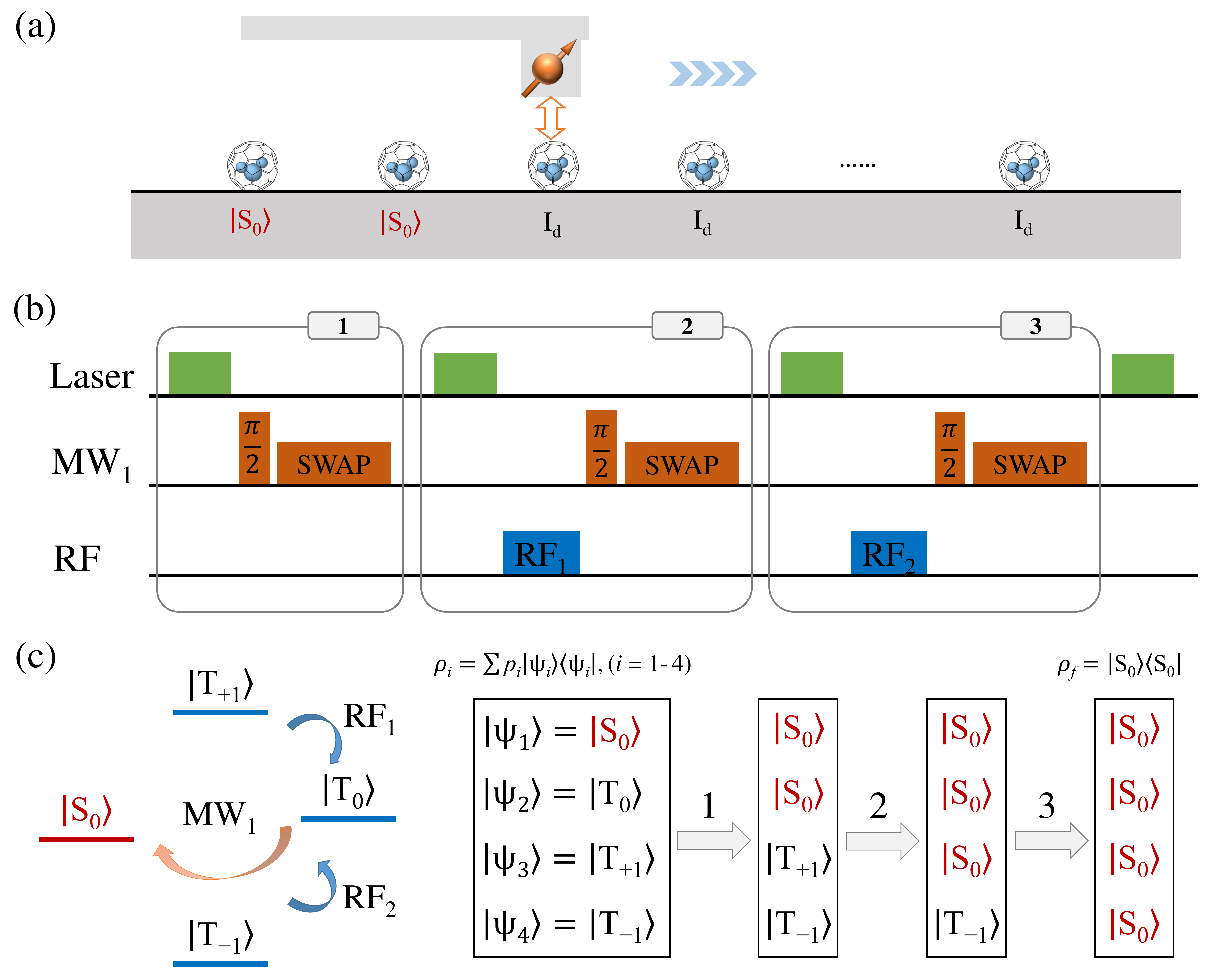}
\caption{Initialization of the DF qubit. (a) The initialization of DF qubits. Qubits are initialized one by one via the quantum actuator.
(b) Pulse sequences applied on the quantum actuator for initialization.
Laser pulses are used to polarize the quantum actuator to the state $|0\rangle$.
The subsequent $\textrm{MW}_1$ $\pi/2$ pulses prepare the actuator on the state $(|0\rangle-|1\rangle)/\sqrt{2}$.
A SWAP gate is applied to initialize the nuclear spin pair states in the DF subspace to state $|S_0\rangle$.
RF pulses, $\textrm{RF}_1$ and $\textrm{RF}_2$, are used to turn the nuclear spin pair's states $|T_{+1}\rangle$ and $|T_{-1}\rangle$ into the DF subspace, respectively.
(c) Evolutions of different nuclear spin pair states under the pulses in (b). DF qubits can be finally initialized to the singlet state $|S_0\rangle$.
\label{fig2}
}
\end{figure}

\textit{Initialization and readout.}
Polarizing nuclear spins usually requires extreme high magnetic fields and low temperature, and reading out nuclear spins directly is also challenging due to their small magnetic moments, especially for nuclear spin pairs used here.
There is only an experiment to sense the coupling of a pair of $^{13}$C in diamond by the NV center \cite{ShiNP}.
The method to obtain the states of a nuclear spin pair still remains elusive.
To overcome this difficulty, we use a quantum actuator as initialization and measurement interfaces for the nuclear-spin pair DF qubit.

For the initialization and readout procedures, the quantum actuator is located above the target DF qubit and other neighboring qubits are about 2~nm away.
Thus the couplings between the quantum actuator and other DF qubits can be ignored (see details in Supplementary Material).
Thus the evolution of the quantum actuator during initialization/readout processes of the target DF qubit will not influence other qubits.
As the Zeeman frequency of the nuclear spin pair under the magnetic field is much larger than the dipole-dipole interactions, the spin-flip process between the inside and outside of the DF subspace can be suppressed.
If required, stronger magnetic field can be adopted in principle.
In the electron spin rotating frame, the Hamiltonian for the system composed of a quantum actuator and a DF qubit is reduced to
\begin{equation} \label{Hlogic}
\begin{split}
H^L= \omega_1(\cos\varphi s_x+\sin\varphi s_y) + AI^{L}_x +2B(s_z+\frac{1}{2})I^L_z,~
\end{split}
\end{equation}
where $\textsl{\textbf{I}}^L$ is the spin-1/2 operator for the logical DF qubit, $\textbf{s}$ is the effective spin-1/2 operator for the quantum actuator.
The first term is the Hamiltonian of the driving field on the quantum actuator, with $\omega_1$ the Rabi frequency and $\varphi$ the phase of resonant driving field.
The second term represents the DF qubit Hamiltonian, with parameter $A$ being the inter-nuclear dipolar interaction strength.
The last one describes the interaction between the quantum actuator and the DF qubit, with parameter $B$ being the difference of the two NV-single nuclear spin couplings (see details in Supplementary Material).

The nuclear spin pairs are approximately a totally mixed state $\rho_i$ at room temperature. Figs.~\ref{fig2}(a)-(c) show the procedures to prepare the spin pair from totally mixed state to the singlet state $|S_0\rangle$.
For the population being initially in the DF subspace, the term $2B(s_z+1/2)I_z^L$ provides the interaction between the quantum actuator and the DF qubit.
Optimal quantum control on the quantum actuator can be used to constitute a SWAP gate between the DF qubit and the actuator.
We take a representative case of $A=-12.7~\textrm{kHz}$ and $B=-6.0~\textrm{kHz}$ as an example (see Ref.~(see details in Supplementary Material) for the geometric arrangement of the spins involved).
The pulse sequences are shown in Fig.~\ref{fig2}(b).
The quantum actuator is polarized with a green laser pulse and is then prepared via applying a $\pi/2$ pulse along $y$ axis to the state $(|1\rangle-|0\rangle)/\sqrt{2}$.
A pulse sequence with adjustable $\varphi$ and $\omega_1$ can be designed to swap the state of the actuator to the DF qubit.
GRadient Ascent Pulse Engineering (GRAPE) algorithm \cite{GRAPE} is used to design the required MW sequence on the actuator.
A 150~$\mu \textrm{s}$ sequence of pulses forming a SWAP gate with fidelity over 99$\%$ with noise effects (see details in Supplementary Material).
After the initialization of the states in the DF subspace, the quantum actuator is polarized to $|0\rangle$ by lasers again to cut off its coupling with the DF qubit so as to protect the DF state $|S_0\rangle$.
The decoherence of DF qubits induced by the pumping process is negligible due to the weak hyperfine couplings between the quantum actuator and the DF qubits \cite{singletrelax, opticaldecoherence}.
For the states of the nuclear spin pair initially out of the DF subspace, resonant RF $\pi$ pulses, $\textrm{RF}_1$ and $\textrm{RF}_2$, are applied to turn the states into the DF subspace (see details in Supplementary Material).
Then the laser and the MW sequence for the initialization of $|S_0\rangle$ are applied (shown as blocks labelled by 2 and 3 in Fig.~\ref{fig2}(b)).
In addition, once the DF qubit is prepared into the singlet state $|S_0\rangle$, it will keep its state even under radio frequency (RF) pulses.
By applying the pulses of these blocks in Fig.~\ref{fig2}(b) sequentially, the DF qubit can be finally prepared to the singlet state $|S_0\rangle$ from a mixed state as shown in Fig.~\ref{fig2}(c).
The fidelity of the whole initialization procedure is estimated over 99$\%$.
The singlet state $|S_0\rangle$ has very long life time, which is even longer that $T_1$ of each nuclear spin \cite{singletrelax}.
Our methods are also appropriate for other different $A$ and $B$ values (see details in Supplementary Material).
Via moving the quantum actuator to various locations of different DF qubits, each DF qubit can be initialized to $|S_0\rangle$ individually with high fidelity.

The readout of DF qubits can also be realized by similar methods as the initialization.
We can also use a SWAP gate to map the DF qubit states onto the actuator. Then the measurement of the quantum actuator, which has already been well solved \cite{NVreadout}, provides the high-fidelity readout of DF qubits.

\begin{figure} 
\centering
\includegraphics[width= 1 \columnwidth]{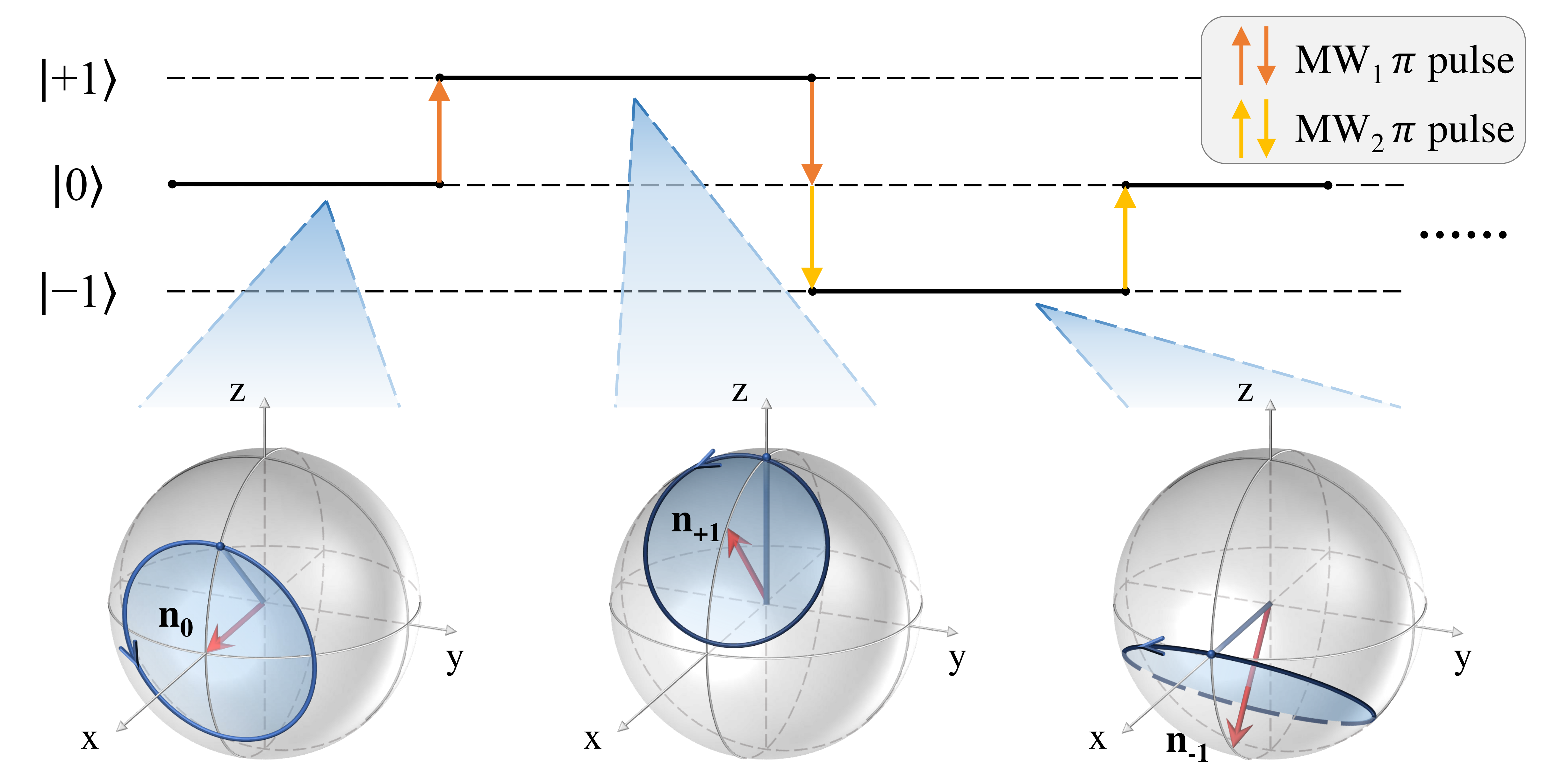}
\caption{Manipulation of the DF qubit.
The quantum actuator is switched among eigenstates $|0, \pm 1\rangle$ by $\textrm{MW}_1$ (orange arrows) and $\textrm{MW}_2$ (yellow arrows) $\pi$ pulses.
Depending on the quantum actuator's states, the state of the DF qubit rotates around distinct axes defined as $\textbf{n}_{0}=\textrm{sign}(A)\hat{x}$ and $\textbf{n}_{\pm1}=A/\Omega_{\pm1}\hat{x} \pm2B/\Omega_{\pm1}\hat{z}$, where $\Omega_{\pm1}=(A^2+4B^2)^{1/2}$. 
\label{fig3}
}
\end{figure}

\textit{Universal quantum gates.}
Couplings between the quantum actuator and DF qubits provide a way to manipulate DF qubits by engineering the actuator.
For the single-qubit manipulation, the quantum actuator is located above the target DF qubit.
The DF qubit will rotate around distinct axes depending on the quantum actuator's eigenstates as shown in Fig.~\ref{fig3}.
When the actuator is polarized to state $|0\rangle$, the coupling term $2B(s_z+1/2)I^L_z$ vanishes and the DF qubit is decoupled from the actuator.
The state of the DF qubit will rotate around $-x$ axis at a speed of $\Omega_0=|A|=12.7~\textrm{kHz}$.
When the quantum actuator is switched onto the states $|\pm1\rangle$, the rotation axes are $\textbf{n}_{\pm1}=A/\Omega_{\pm1}\hat{x} \pm2B/\Omega_{\pm1}\hat{z}$, where $\Omega_{\pm1}=(A^2+4B^2)^{1/2}=17.5~\textrm{kHz}$ is the speed.
This process is limited by the $T_1$ relaxation time of the quantum actuator, which can reach several milliseconds even at room temperature.
Rotations along two noncommutable axes construct a universal set of single-qubit gate \cite{NielsenQCQI}.
And we can time-optimally control the qubit by designing an appropriate sequence of actuator's states \cite{actuatorcontrol}.

\begin{figure} 
\centering
\includegraphics[width=1 \columnwidth]{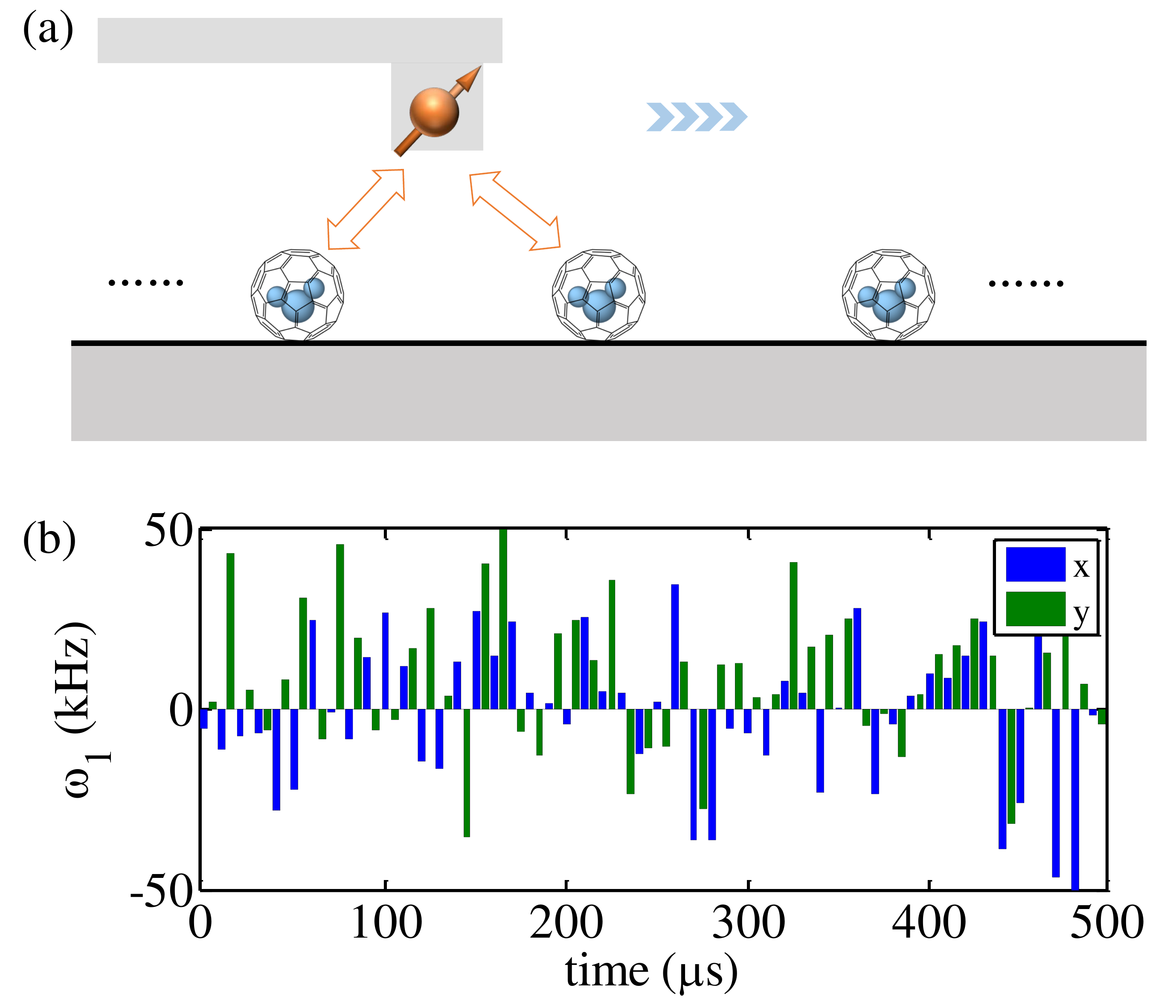}
\caption{CNOT gate between DF qubits.
(a) The quantum actuator locates between two qubits, so as to realize a CNOT gate between the DF qubits.
(b) The designed pulse sequence for the CNOT gate. The pulses are applied on the quantum actuator along $x$ axis (blue) and $y$ axis (green) alternately. The gate fidelity is higher than 99$\%$ even with noise effects. 
\label{fig4}
}
\end{figure}

Two-qubit gates can also be realized with the quantum actuator.
When the actuator locates between two DF qubits, it will couple with both of the DF qubits (Fig.~\ref{fig4}(a)).
Optimal quantum control can be used to construct two-qubit gates \cite{tolerantgate}.
Through applying on the quantum actuator a sequence of MW pulses obtained from the GRAPE algorithm, a two-qubit operation, i.e., the CNOT gate between two DF qubits, can be achieved.
Thus the universal set of quantum gates can be realized by combining single-qubit and CNOT gates.
Two neighbouring DF qubits with parameters $A_1=-12.7~\textrm{kHz}$, $B_1=8.0~\textrm{kHz}$, $A_2=-5.2~\textrm{kHz}$ and $B_2=-3.7~\textrm{kHz}$ are considered as an instance in the simulation (see details in Supplementary Material).
The pulse sequence shown in Fig.~\ref{fig4}(b) is an example obtained via the GRAPE algorithm.
The pulses are along $x$ and $y$ axes alternately and the total time of the sequence is 500~$\mu$s with each MW pulse of 5~$\mu$s.
The CNOT gate fidelity is higher than 99$\%$ even with noise effects (see details in Supplementary Material).
Here, typical mechanical speed of moving tips (quantum actuator) can be around 0.23 nm/$\mu$s \cite{AFM speed}. Thus the cost time of moving actuator can be neglected comparing with the quantum gates.
With single qubit gates and two-qubit CNOT gates, universal quantum gates can be realized \cite{quantumactuator}.

\textit{Scalability}.
In our architecture, the array of well-prepared DF qubits spans a large Hilbert space with a dimension of $O(2^N)$, where $N$ is the number of qubits.
A universal set of quantum gates can be applied on the DF qubits under the help of the quantum actuator with high fidelities exceeding the threshold required for a fault-tolerant quantum information processing implementation \cite{faultthreshold}.
Quantum information can be transferred throughout the system by sequentially moving the quantum actuator between neighboring DF qubits.
To connect any two qubits requires no more than $O(N)$ pairwise operations intermediated by the quantum actuator.
Arbitrary quantum logic transformations can be performed on qubits in a pairwise form.
As a result, any desired quantum logic circuit of $n$ logic gates can be built up in $O(nN)$ operations \cite{quantumactuator}.


In summary, we have put forward a scalable architecture for quantum computation based on quantum actuated nuclear-spin DF qubits.
By utilizing the NV electron spin in diamond as the quantum actuator, initialization, readout and manipulation of DF qubits can be realized.
In addition, DF qubits are immune to collective noise, allowing the information to survive for a period extended by orders of magnitudes compared to that of nuclear spins.
The scalability of the computation is also discussed.
Our results provide a promising candidate for large-scale quantum computation within the reach of potential experimental technologies.

We would like to thank V. Vedral and M. Gu for helpful discussions.
This work was supported by the 973 Program (Grant No.\ 2013CB921800 and No.\ 2016YFB0501603), the NNSFC (Grant Nos. 11227901, 31470835, 21233007, 21303175, 21322305, 11374305 and 11274299), the ``Strategic Priority Research Program (B)'' of the CAS (Grant Nos. XDB01030400 and 01020000). F.S. and X.R. thank the Youth Innovation
Promotion Association of Chinese Academy of Sciences for their support.



\section{Supplementary material}

\section {I. Hamiltonian simplification}
The total Hamiltonian of our architecture composed of a quantum actuator and nuclear spin pairs is described as
\begin{equation}
\begin{split}
H_0= H_{\textrm{NV}}+H_{\textrm{pairs}}+H_{\textrm{NV}-\textrm{pairs}}.
\end{split}
\end{equation}

The Hamiltonian of the quantum actuator is given by
\begin{equation}
H_{\textrm{NV}}=DS^2_z+\omega_eS_z,
\end{equation}
where $\textbf{S}$ is the spin-1 operator, $\omega_e$ is the Zeeman frequency corresponding to the static magnetic field $B_0$, and $D = 2.87~\textrm{GHz}$ is the zero-field splitting of the NV center.

The Hamiltonian of the nuclear spin pairs is written as
\begin{equation}
H_{\textrm{pairs}}=\sum\limits _m (\omega_iI_z^{m,p}+\omega_iI_z^{m,q}+\textsl{\textbf{I}}^{m,p} \cdot \textsl{\textbf{D}}^{m,pq} \cdot \textsl{\textbf{I}}^{m,q}),
\end{equation}
where $\textbf{I}^{m,p}$ and $\textbf{I}^{m,q}$ are spin-1/2 operators for nuclear spins in $m$th spin pairs, $\omega_i$ is the nuclear spin Zeeman frequency, and $\textbf{D}^{m,pq}$ is their interaction tensor between spin $P$ and $Q$.
The couplings between different nuclear spin pairs are small enough (at Hertz level) to be neglected.

The interactions between the quantum actuator and nuclear spin pairs are
\begin{equation}
H_{\textrm{NV-pairs}}=\sum\limits_m(\textsl{\textbf{S}}\cdot \textsl{\textbf{D}}^{m,p} \cdot \textsl{\textbf{I}}^{m,p} +\textsl{\textbf{S}}\cdot \textsl{\textbf{D}}^{m,q} \cdot \textsl{\textbf{I}}^{m,q}),
\end{equation}
where $\textbf{D}^{m, p(q)}$ is the interaction tensor between the quantum actuator and the nuclear spin $P(Q)$ in $m$th spin pairs.

For the process of initialization, readout and single-qubit control, the quantum actuator locates above the target DF qubit with other nuclear spin pairs being ignored because of the weak couplings (see details in Part II).
The Hamiltonian of the system composed of the quantum actuator and one pair of nuclear spins ($m=1$) is 
\begin{equation}
\begin{split}
H_{\textrm{pair}}=\omega_iI_z^p+\omega_iI_z^q+\textsl{\textbf{I}}^p \cdot \textsl{\textbf{D}}^{pq} \cdot \textsl{\textbf{I}}^q,\\
H_{\textrm{NV-pair}}=\textsl{\textbf{S}}\cdot \textsl{\textbf{D}}^{p} \cdot \textsl{\textbf{I}}^p +\textsl{\textbf{S}}\cdot \textsl{\textbf{D}}^{q} \cdot \textsl{\textbf{I}}^q.
\end{split}
\end{equation}
As the large zero-field splitting allows for the secular approximation, the effective Hamiltonian of the interactions can be rewritten as
\begin{equation}\label{Hamirot}
\begin{split}
H_{\textrm{NV-pair}}&= \sum_{j=x,y,z}(D_{zj}^pS_zI_j^p +D_{zj}^qS_zI_j^q).
\end{split}
\end{equation}
Here we take the dipole-dipole interaction as an example.
The Zeeman splitting $\omega_i$ of nuclear spins under the magnetic field is much larger than the interactions, i.e., $\omega_i>D_{ij}^{pq}$, $D_{ij}^{p(q)}$.
If required, stronger magnetic field can be adopted in principle.
The polarized nuclear spin states $|T_{+1}\rangle=\mid\uparrow\uparrow\rangle$ and $|T_{-1}\rangle=\mid\downarrow\downarrow\rangle$ are energetically separated from the unpolarized states $|T_0\rangle=(\mid\uparrow\downarrow\rangle + \mid\downarrow\uparrow\rangle)/\sqrt{\textrm{2}}$ and $|S_0\rangle=(\mid\uparrow\downarrow\rangle -\mid\downarrow\uparrow\rangle)/\sqrt{\textrm{2}}$ which span a DF subspace.
There is no leakage between the inside and outside of the DF subspace.
The Hamiltonians of the DF qubit and its coupling with the quantum actuator are given by
\begin{equation}\label{Hlogic1}
\begin{split}
H_{\textrm{DF}}&=\frac{1}{2}(D^{pq}_{xx}+D^{pq}_{yy})I^{L}_x-\frac{1}{4}D^{pq}_{zz},\\
H_{\textrm{NV-DF}}&=\frac{1}{2}(D^{pq}_{yx}-D^{pq}_{xy})S_xI^{L}_y +(D^p_{zz}-D^q_{zz})S_z I^{L}_z,
\end{split}
\end{equation}
where $\textbf{I}^L$ is the spin-1/2 operator for the logical DF qubit, defined as
\begin{equation}
\begin{split}
I^{L}_x&=\frac{1}{2}(\mid\uparrow\downarrow\rangle\langle\downarrow\uparrow\mid+\mid\downarrow\uparrow\rangle\langle\uparrow\downarrow\mid),\\ I^{L}_y&=\frac{1}{2}(-i\mid\uparrow\downarrow\rangle\langle\downarrow\uparrow\mid+i\mid\downarrow\uparrow\rangle\langle\uparrow\downarrow\mid),\\
I^{L}_z&=\frac{1}{2}(\mid\uparrow\downarrow\rangle\langle\uparrow\downarrow\mid-\mid\downarrow\uparrow\rangle\langle\downarrow\uparrow\mid).
\end{split}
\end{equation}
As $D^{pq}_{yx}$ equals to $D^{pq}_{xy}$, and $\frac{1}{4}D^{pq}_{zz}$ only induces overall phase, the Hamiltonian of NV-DF system can be reduced to
\begin{equation}\label{Hlogic2}
\begin{split}
H_0^{L}=DS^2_z+\omega_eS_z+AI^{L}_x +2BS_zI^L_z,
\end{split}
\end{equation}
where $A=\frac{1}{2}(D^{pq}_{xx}+D^{pq}_{yy})$ and $B=\frac{1}{2}(D^{p}_{zz}-D^{q}_{zz})$.

When the quantum actuator is driven by the microwave (MW) pulses on resonance with transition $|0\rangle\leftrightarrow |\textrm{+}1\rangle$ (or alternatively $|0\rangle\leftrightarrow |\textrm{-}1\rangle$), it is reduced to an effective two-level system. 
In the electron spin rotating frame, the Hamiltonian of the quantum actuator with the control field is
\begin{equation}\label{Hc}
H_\textrm{NV}^\textrm{eff} =\omega_1(\cos\varphi s_x+\sin\varphi s_y),
\end{equation}
where $\omega_1$ is the Rabi frequency and $\varphi$ is the phase of resonant driving field.
$\textbf{s}$ is the effective spin-1/2 operator for the quantum actuator, defined as
\begin{equation}
\begin{split}
s_x&=\frac{1}{2}(|\textrm{+}1\rangle\langle0|+|0\rangle\langle\textrm{+}1|),\\ s_y&=\frac{1}{2}(-i|\textrm{+}1\rangle\langle0|+i|0\rangle\langle\textrm{+}1|),\\
s_z&=\frac{1}{2}(|\textrm{+}1\rangle\langle\textrm{+}1|-|0\rangle\langle0|).
\end{split}
\end{equation}
In the rotating frame, the Hamiltonian of NV-DF system under the control field takes the form
\begin{equation}\label{Hlogic3}
\begin{split}
H^L=\omega_1(\cos\varphi s_x+\sin\varphi s_y)+AI^{L}_x +2B(s_z+\frac{1}{2})I^L_z.
\end{split}
\end{equation}

\section{II. Estimation of coupling strength}
For the process to handle with one DF qubit, such as the initialization and readout, the quantum actuator is located above the target DF qubit (qubit-\textit{i}) and far away from the others (Fig.~\ref{fig1}(a)). 
\begin{figure} [http]
\centering
\includegraphics[width= \columnwidth]{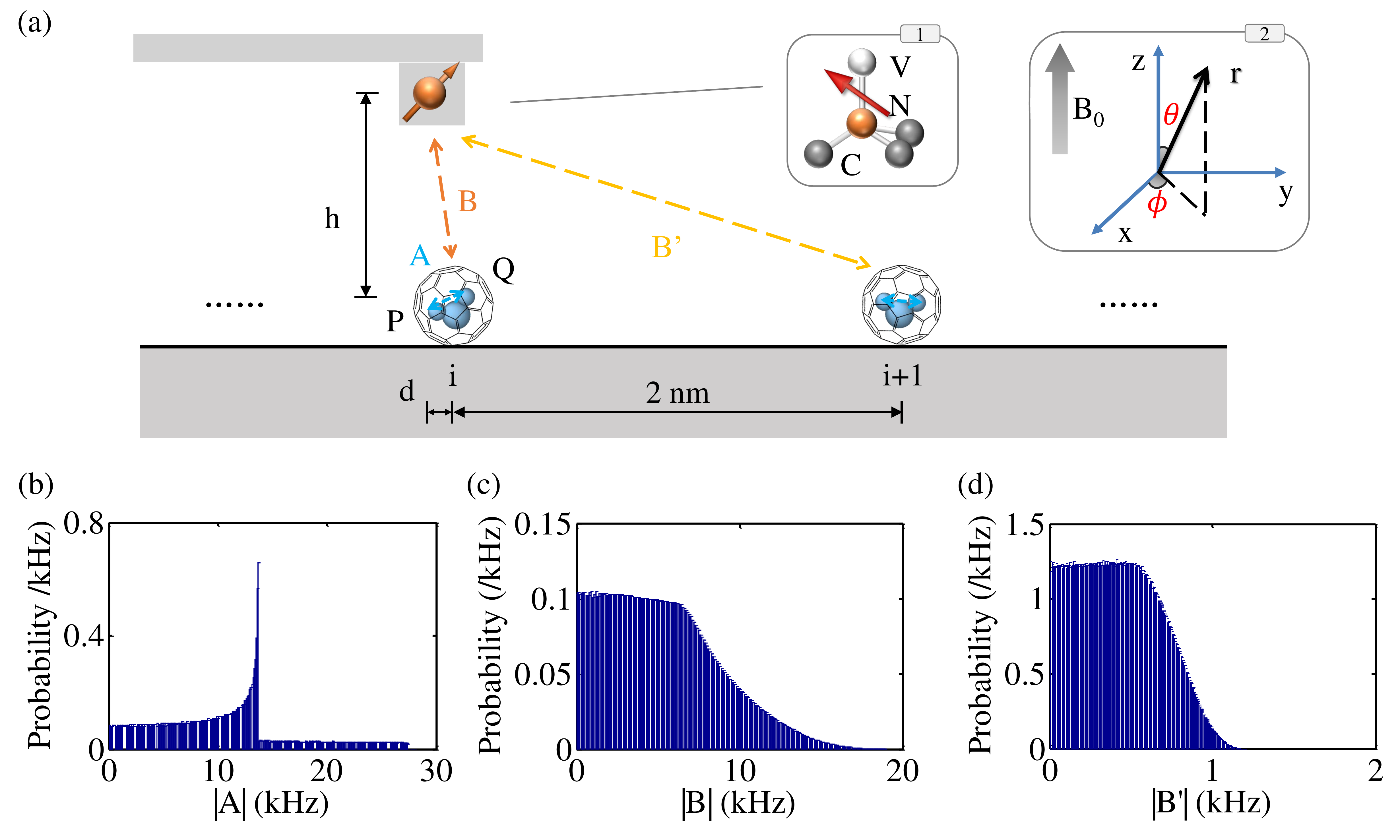}
\caption{Initialization of the DF qubit.
(a) The quantum actuator initializes the target DF qubit $i$.
The internal nuclear spin coupling are denoted as $A$ (blue).
The quantum actuator interacts with the target DF qubit with coupling strength $B$ (orange), and with the neighboring DF quibt with strength $B'$ (yellow).
$h$ and $d$ represent the vertical and horizontal distance between the quantum actuator and the midpoint of the nuclear spin pair.
Inset labelled by 1 shows the atomic structure.
Inset labelled by 2 shows the coordinate system.
$z$ axis is along the N-V symmetry axis ([1 1 1] crystal axis), which is also perpendicular to base surface. $x$ and $y$ axes are along the direction of the DF qubit array.
Vector $\textbf{r}$ represents the alignment of two spins.
$\theta$ is the angle between vector and $z$ axis, and $\phi$ represents the angle between the projection in $xy$ plane and $x$ axis.
(b)-(d) Distribution of coupling strength $|A|$ (b), $|B|$ (c) and $|B'|$ (d). 
\label{fig1}}
\end{figure}
The coordinate system is defined as shown in the inset labelled by 2 in Fig.~\ref{fig1}(a).
$z$ axis is set along the N-V symmetry axis ([1 1 1] crystal axis), which is also perpendicular to base surface for convenience.
$x$ and $y$ axes are along the directions of DF qubit array.
The midpoint of the nuclear spin pair is set as the origin.
$\theta$ represents the angle between vector $\textbf{r}$ and $z$ axis, and $\phi$ is the angle between the projection in $xy$ plane and $x$ axis.

The inter-nuclear dipolar interaction strength $A=\frac{1}{2}(D^{pq}_{xx}+D^{pq}_{yy})$ is dependent on the relative position of nuclear spin $P$ and $Q$.
As the distance between two protons in H$_2$O is fixed, 0.1635~nm, the angles $\theta$ and $\phi$ are the variables.
Averaging the orientation of the nuclear spin pair, distribution of $|A|$ can be calculated as shown in Fig.~\ref{fig1}(b).
The most probability of values locates around 14~kHz, which we used in the simulations.

The NV-DF interaction $B =\frac{1}{2}(D^{p}_{zz}-D^{q}_{zz})$ is the difference of the two NV-single nuclear spin couplings and is relevant with three spins' positions.
The distribution of $|B|$ can be calculated as shown in Fig.~\ref{fig1}(c) with the actuator locating in the range of $1.2<h<1.5~\textrm{nm}$ and $d<0.4$~nm, and $\textbf{r}_{pq}$ being along arbitrary direction.
The value of -6.0~kHz used in the simulation is a representative case.

The couplings between the quantum actuator and other DF qubits are unwanted in the initialization process.
Here we estimate the coupling strength between the quantum actuator and the nearest neighboring DF qubit (2~nm away), denoted as $B'$, with results in Fig.~\ref{fig1}(d).
When the value of $B$ locates at a common position, e.g. 6~kHz, the maximum of $B'$ is smaller than one fifth of $B$.
And we can even adjust the position of quantum actuator with the smallest possible valve of $B'$.
Other remote DF qubits can also be neglected because coupling strength decays rapidly with the increase of distance.

\section{III. MW pulse sequence}
In the initialization procedure, a sequence of pulses is applied on the quantum actuator to construct a SWAP gate between the actuator and the DF qubit.
The geometric arrangements of the spins simulated in the main text are described as follows.
$r_{pq}=0.1635~\textrm{nm}$, $\theta_{pq}=0.45\pi$, $\phi_{pq}=0$, and $r_{p-nv} = 1.3~\textrm{nm}$, $\theta_{p-nv} =0.05\pi$, $\phi_{p-nv} = 0.2\pi$.
A concrete MW pulse sequence calculated by GRAPE is shown in Fig.~\ref{fig2}(a).
The pulses are applied along $x$ and $y$ alternately with various Rabi frequencies.
The total time is 150~$\mu \textrm{s}$ and there are 100 pulses in the sequence.
The fidelity of the SWAP gate is 99.9$\%$ without noises.
In practice, inhomogeneity of $\omega_1$, denoted as $\delta_1$, is one of the main noises in experiments.
It mainly comes from the control field error because of the static fluctuation of the microwave power and the distortion of the pulse owing to the finite bandwidth.
Our pulse sequence can resist to this noise as shown in Fig.~\ref{fig2}(a).

Additionally, mechanical vibration of tips (quantum actuator's position) is also a source of noise.
Here, mechanical stability can reach 10 pm or even smaller \cite{AFM}, causing a variation at the level of 1$\%$ of the coupling between the actuator and DF qubits.
For example, 10 pm vibration of $r_{p-nv}$ (originally equal to 1.3 nm) causes a coupling variation of 0.2~kHz (3$\%$ of $B$).
With such a error, the SWAP gate fidelity can still reach 99.4$\%$.


\begin{figure} [http]
\centering
\includegraphics[width= \columnwidth]{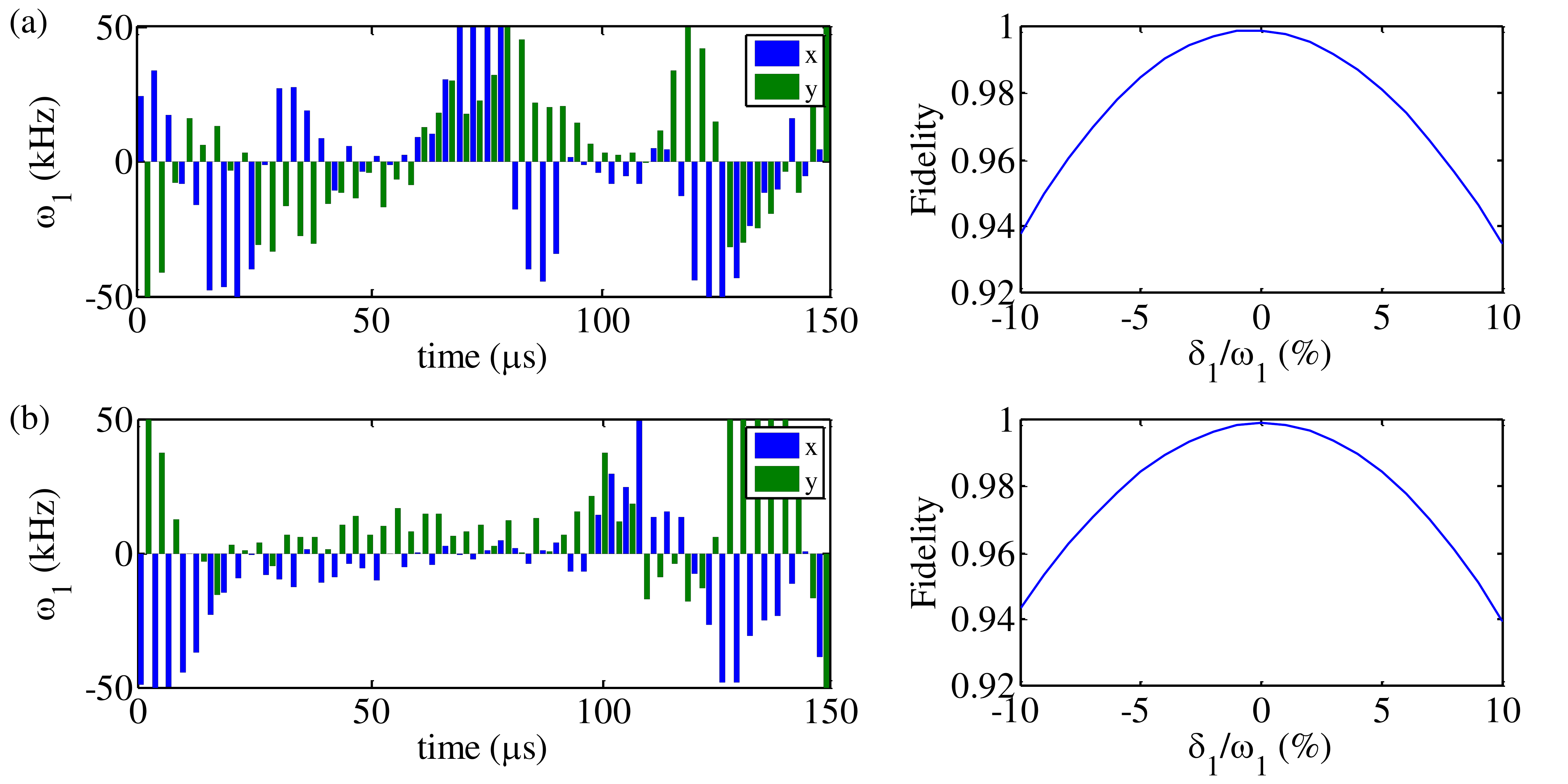}
\caption{Pulse sequences for the SWAP gate.
(a) The designed pulse sequence for the SWAP gate between the quantum actuator and the DF qubit with the case of $A=-12.7~\textrm{kHz}$ and $B=-6.0~\textrm{kHz}$ (left). The alternant pulses along $x$ (blue) and $y$ (green) axes are applied on resonance of $|0\rangle\leftrightarrow |1\rangle$ of the quantum actuator. The total time is $150~\mu \textrm{s}$ and each pulse is $1.5~\mu \textrm{s}$.
The pulse sequence can resist to the noise $\delta_1$ (right).
(b) The designed pulse sequence for the SWAP gate with the case of $A= -5.2~\textrm{kHz}$ and $B= -6.7~\textrm{kHz}$ (left) and its robustness to $\delta_1$ (right).
\label{fig2}}
\end{figure}

This method is also appropriate for other values of $A$ and $B$.
For the DF qubit with $A=-5.2~\textrm{kHz}$ and $B=-6.7~\textrm{kHz}$.
The corresponding positions are $r_{pq}= 0.1635~\textrm{nm}$, $\theta_{pq}=0.35\pi$. $\phi_{pq}=0$. $r_{p-nv} = 1.4~\textrm{nm}$, $\theta_{p-nv} =0.07\pi$, $\phi_{p-nv} = 0.1\pi$.
In Fig.~\ref{fig3}(b), an example is presented, a SWAP gate between the quantum actuator and the DF qubit with high fidelity.


\section{IV. RF pulses}
For the states of the nuclear spin pair initially out of the DF subspace, resonant radio frequency (RF) $\pi$ pulses, $\textrm{RF}_1$ and $\textrm{RF}_2$, are applied to turn them into the DF subspace.
The Hamiltonian of the RF pulses is
\begin{equation}\label{RF}
H_{\textrm{RF},k} =\frac{\sqrt2}{2}\omega_{\textrm{RF}}\cos(2\pi\nu_kt) (I_x^p+I_x^q),\,k=1\,\textrm{or}\, 2,
\end{equation}
where $k$ represents the RF pulses with different frequency $\nu_k$, and $\omega_{\textrm{RF}}$ is the Rabi strength.
In the bases of the eigenstates of nuclear spin pair, $|T_{+1}\rangle$, $|T_{-1}\rangle$, $|T_{0}\rangle$ and $|S_{0}\rangle$, $H_{\textrm{RF},k}$ can be rewritten as
\begin{equation}\label{RF}
\begin{split}
H_{\textrm{RF},k} =\omega_{\textrm{RF}}\cos(2\pi\nu_kt)(|T_{+1}\rangle\langle T_{0}|+|T_{0}\rangle\langle T_{+1}|\\
+|T_{-1}\rangle\langle T_{0}|+|T_{0}\rangle\langle T_{-1}|)/2,\,k=1\, \textrm{or}\,2.
\end{split}
\end{equation}
RF pulses can cause the transitions between $|T_{+1}\rangle$, $|T_{-1}\rangle$ and $|T_{0}\rangle$.
We denote the energy splitting between $|T_{\pm1}\rangle$ and $|T_{0}\rangle$ as $\Delta_{1,2}$. 
The difference of $\Delta_1$ and $\Delta_2$ equals three times of $A$, about 40~kHz here.
$\nu_1$ is set equal to $\Delta_1$, and $\omega_{\textrm{RF}}=5~\textrm{kHz}$ is much smaller than $3A$.
Then a $\textrm{RF}_1$ $\pi$ pulse costing 100~$\mu$s will turn the state $|T_{+1}\rangle$ to $|T_{0}\rangle$ without flips between $|T_{-1}\rangle$ and $|T_{0}\rangle$.
The fidelity is over $99\%$.
$\textrm{RF}_2$ is set similarly with $\nu_2=\Delta_2$ to deal with the initial state $|T_{-1}\rangle$.

\begin{figure} [http]
\centering
\includegraphics[width= \columnwidth]{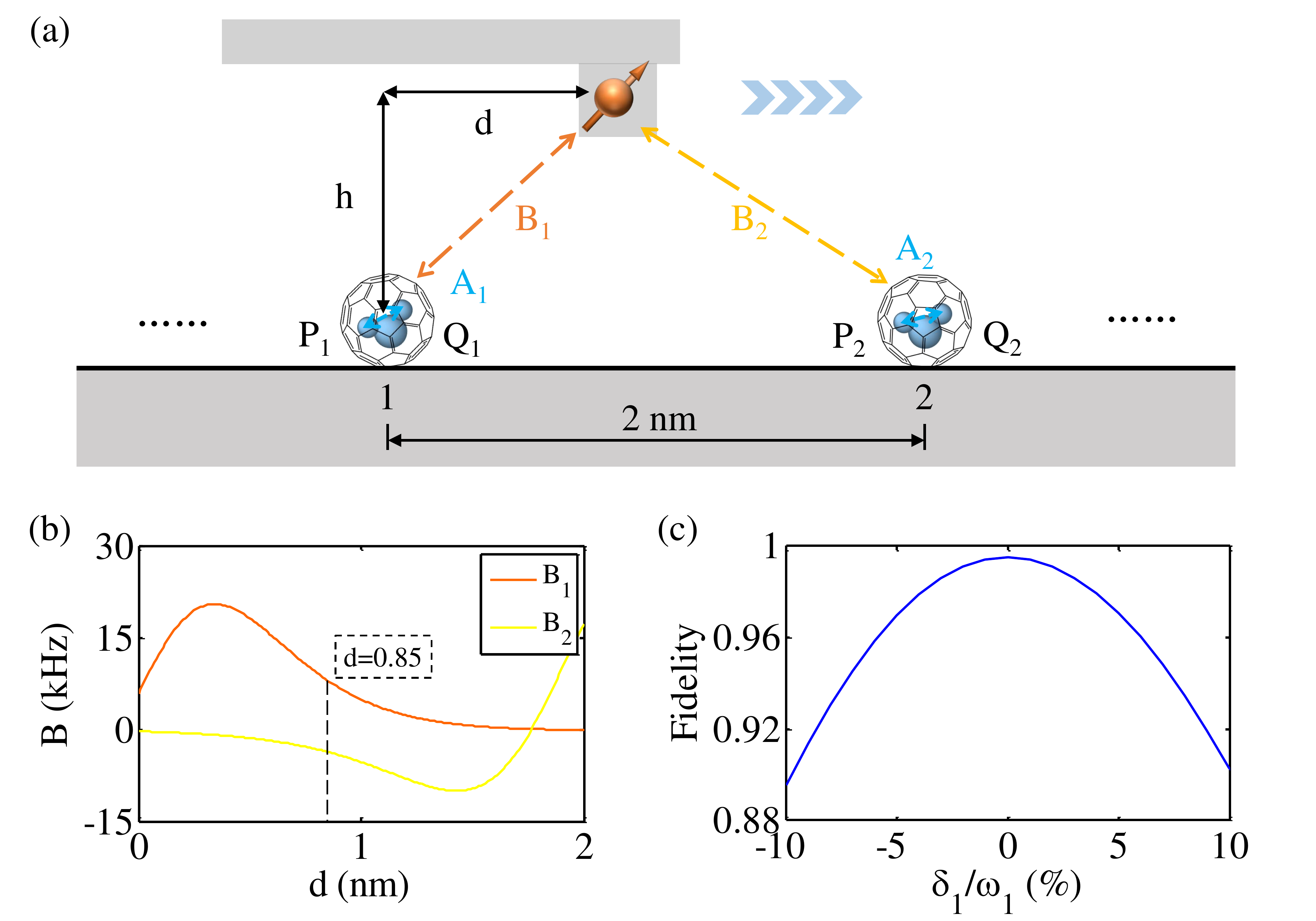}
\caption{The CNOT gate between DF qubits.
(a) The quantum actuator locates between two qubits, qubit-1 and qubit-2, with the corresponding coupling strength $A_1$, $B_1$ and $A_2$, $B_2$, so as to realize a CNOT gate between the DF qubits.
(b) $B_1$ and $B_2$ serve as functions of the displacement $d$.
The distance of 0.85~nm is taken as the example simulated in the main text with $B_1=8.0~\textrm{kHz}$ and $B_2=-3.7~\textrm{kHz}$.
(\textbf{c}) The CNOT gate obtained in the main text can resist to $\delta_1$ noise, and the maximum of the fidelity is 0.995.
\label{fig3}}
\end{figure}

\section{V. Quantum gates}

For the process to construct a two-qubit gate between DF qubits, the quantum actuator is located between two qubits in order to achieve strong interactions with both of them (Fig.~\ref{fig3}(a))
The nuclear spins in two pairs are denoted as $P_1$, $Q_1$, $P_2$ and $Q_2$.
The coupling between the quantum actuator and qubit-1 (qubit-2) is denoted as $B_1$ ($B_2$).
In the main text, we simulated a case of $A_1=-12.7~\textrm{kHz}$ and $A_2=-5.2~\textrm{kHz}$ as an instance.

$A$ is just related to the relative positions of the nuclear spins.
The position parameters of the two nuclear spin pairs are $r_{pq1}=0.1635~\textrm{nm}$, $\theta_{pq1}=0.45\pi$, $\phi_{pq1}=0$ and $r_{pq2}=0.1635~\textrm{nm}$, $\theta_{pq2}=0.35\pi$, $\phi_{pq2}=0$, respectively.
The distance between two pairs is 2~nm.
the vertical height $h$ of the quantum actuator is set 1~nm.
When the quantum actuator moves from left to right, the couplings with the DF qubits, $B_1$ and $B_2$, will vary accordingly.
Fig.~\ref{fig3}(b) shows the changing curves of $B_1$ and $B_2$ as a function of displacement $d$.
When the actuator locates near the mid-position, both the couplings are considerable, which is appropriate to construct a two-qubit gate between DF qubits.
We take the quantum actuator with $d= 0.85~\textrm{nm}$ as the example, in which $B_1=8.0~\textrm{kHz}$ and $B_2=-3.7~\textrm{kHz}$.
The CNOT gate obtained in the main text can also resist to the noise $\delta_1$ as shown in Fig.~\ref{fig3}(c).

Then we consider the influence of mechanical instability on quantum gates.
The mechanical stability can reach 10 pm or even smaller \cite{AFM}, which causes a variation at the level of 1$\%$ of the coupling between the actuator and DF qubits.
For example, 10 pm vibration of $r_{p-nv}$ causes a coupling variation of 0.2~kHz (3$\%$ of $B$), which will slightly vary the evolution Hamiltonian of the DF qubit, $AI^{L}_x +2BS_zI^L_z$.
In the simulation, with such an instability, the deviation of the qubit's evolution is below 0.09$\%$.
So the influence of mechanical instability on quantum gates can be ignored.

\section{Supplementary reference}

\end{document}